\documentclass[preprint,12pt]{elsarticle}

\usepackage{amssymb}
\usepackage{amsmath}

\journal{Journal of Theoretical Biology}

\begin{document}

\begin{frontmatter}



\title{Why aphids are not pests in cacao? An approach based on a
  predator-prey model with aging}


\author[dfisuefs]{Vladimir R. V. Assis} 
\author[dfisuefs]{Nazareno G. F. Medeiros} 
\affiliation[dfisuefs]{
  organization={Departamento de F{\'\i}sica, Universidade Estadual de
    Feira de Santana},
  addressline={Av. Transnordestina S/N, Novo Horizonte}, 
  city={Feira de Santana},
  postcode={44031-460}, 
  state={Bahia},
  country={Brazil}}

\author[dcbiouefs]{Evandro N. Silva} 
\affiliation[dcbiouefs]{
  organization={Departamento de Ci\^encias Biol\'ogicas, Universidade Estadual de
    Feira de Santana},
  addressline={Av. Transnordestina S/N, Novo Horizonte}, 
  city={Feira de Santana},
  postcode={44031-460}, 
  state={Bahia},
  country={Brazil}}

\author[dcnmeufscar]{Alexandre Colato} 
\affiliation[dcnmeufscar]{ 
  organization={Departamento de Ci\^encias da Natureza, Matem\'atica e
    Educa\c{c}\~ao, Universidade Federal de S\~ao Carlos},
  addressline={Rodovia Anhanguera, Km 174}, 
  city={Araras},
  postcode={13600-970}, 
  state={S\~ao Paulo},
  country={Brazil}}

\author[dfisuefs,ppgmuefs]{Ana T. C. Silva} 
\affiliation[ppgmuefs]{
  organization={PPGM, Universidade Estadual de
    Feira de Santana},
  addressline={Av. Transnordestina S/N, Novo Horizonte}, 
  city={Feira de Santana},
  postcode={44031-460}, 
  state={Bahia},
  country={Brazil}}

\begin{abstract}
  We studied a mean-field predator-prey model with aging to simulate
  the \mbox{interaction} between aphids (\textit{Toxoptera aurantii})
  and syrphid larvae in \mbox{cacao} farms in Ilheus, Bahia. Based on
  the classical predator-prey model, we \mbox{propose} a system of
  differential equations with three rate equations. \mbox{Unlike} the
  original Lotka-Volterra model, our model includes two aphid
  population classes: juveniles (non-breeding) and adult females
  (asexually breeding). We obtained steady-state solutions for
  juvenile and adult populations by \mbox{analyzing} the stability of
  the fixed points as a function of model \mbox{parameters}. The
  results show that the absorbing state (zero prey population) is
  always possible, but not consistently stable. A nonzero stationary
  solution is achievable with appropriate parameter values. Using
  phase diagrams, we analyzed the \mbox{stationary} solution,
  providing a comprehensive understanding of the \mbox{dynamics}
  involved. Simulations on complete graphs yielded \mbox{results}
  closely matching the differential equations. We also
  \mbox{performed} simulations on \mbox{random} networks to highlight
  the influence of \mbox{network} topology on \mbox{system}
  behavior. Our findings highlight the critical role of life-stage
  structure, \mbox{predation}, and spatial variation in stabilizing
  predator-prey \mbox{systems}. This emphasizes the importance of
  network effects in population dynamics and refines the framework for
  biological pest control in agriculture. Ultimately, our research
  contributes to sustainable agricultural practices.
\end{abstract}



\begin{keyword}


  
  aphids \sep cacao cultivation \sep population dynamics \sep mean-field
approximation 
  
\end{keyword}

\end{frontmatter}



\section{Introduction}
\label{intro}

Studying predator-prey dynamics is essential for understanding
ecological interactions and population control mechanisms in
agricultural ecosystems~\mbox{\cite{Adis1984, Armbrecht2001}}. Various
factors influence these dynamics, including environmental conditions,
species interactions, and network structures~\cite{Dejean2003}. In
addition to these factors, the role of natural enemies in influencing
the population dynamics of tropical insects has been
well-documented~\cite{Perfecto1997, VandermeerPerfecto2006}. For
instance, Godfray and Hassell (1987) argue that natural enemies may be
a cause of discrete generations in tropical
insects~\cite{GodfrayHassell1987}. Moreover, recent studies have
\mbox{examined}
cacao plants' interactions between aphid predators and ant-aphid
mutualism~\cite{Canedo-Junior2019,Nelson2019}. Silva and Perfecto
(2013) explore how these \mbox{interactions} contribute to the
coexistence of aphid predators, suggesting that ant-aphid mutualism
may play a significant role in shaping predator-prey
dynamics~\cite{SilvaPerfecto2013}.

The relevance of classical predator-prey models to real-world
scenarios has been questioned \cite{Ginzburg1992}. Dost\'alkov\'a,
Kindlmann, and Dixon (2002) \mbox{discuss} the applicability of these
models, highlighting their limitations and the need for incorporating
more realistic ecological interactions
\cite{DostalkovaKindlmannDixon2002}. Farina and \linebreak
\mbox{Dennunzio}~(2008) extend this discussion by developing a
predator-prey cellular automaton that includes parasitic interactions
and environmental effects, providing a more nuanced understanding of
these \mbox{dynamics}~\cite{FarinaDennunzio2008}. Furthermore,
Stewart, Logsdon, and Kelley (2005) conducted an empirical study on
the evolution of virulence under both horizontal and vertical
transmission, providing insights into how these transmission modes
influence pathogen \mbox{dynamics} and
virulence~\cite{StewartLogsdonKelley2005}. Aphids are among the most
serious pests, causing severe economic losses in several crops
worldwide~\cite{VanEmden2007}. In cacao, however, the aphid
\mbox{\textit{Toxoptera aurantii}} is not perceived as a pest due to
effective biological \mbox{control} by insect
predators~\cite{Entwistle1972}. In cacao crops, the interactions
between aphids and predators, mainly ladybugs, hoverflies, and
lacewings, are poorly known~\cite{Labrie2023}.

One of the reasons why aphids cause low damage to cacao plants is the
effectiveness of an assemblage of predators, primarily species of the
families Coccinellidae (Coleoptera), Syrphidae (Diptera), Chrysopidae
and \mbox{Hemerobiidae} (Neuroptera)~\cite{Dixon2000}. In Ghana,
Firempong and Kumar (1975) verified in field experiments that
hoverflies and ladybug beetles were able to effectively suppress
\textit{T. aurantii} in cacao due to three factors: the \linebreak
\mbox{synchronization} of predator and prey colonization, the ability
of predators to increase prey consumption at higher prey densities
(functional response) and the complementary effect of the different
coexisting species of natural enemies~\cite{FirempongKumar1975}. The
effectiveness of biological control in various agricultural systems
has been well-documented~\cite{VanDriesche1996}. Carver (1978)
highlighted the role of natural predators in controlling pest
populations in several crops \cite{Carver1978}. \mbox{Numerous}
studies have highlighted the complex interactions within agricultural
ecosystems~\cite{Altieri1999, Tilman2002, Power2010}. Krebs (2001)
provided a comprehensive overview of ecological methodologies and
their applications in understanding these interactions
\cite{Krebs2001}.

The dynamics of natural regulation of insect pests through biological
control is an important achievement for developing sustainable,
ecologically based farming practices \cite{Altieri1995}. Modelling the
dynamics of species interactions can be
paramount to reaching such a level of understanding, as demonstrated
for a natural-scale enemie interaction in shade coffee in Mexico
\cite{VandermeerPerfecto2006}. Studying predator-prey interactions is
fundamental to ecology, particularly in agroecosystems where pests can
significantly impact crop yield and quality \cite{Barbosa1998}. These
environments are complex, with various species interacting, often
resulting in intricate population dynamics \cite{Gurr2003}. These
interactions can determine the success or failure of crop production,
making it crucial to \mbox{comprehend} the underlying mechanisms that
govern these \mbox{relationships}~\cite{Power2010}. In many
agricultural systems, certain insects can become pests, causing
\mbox{extensive} crop damage~\cite{Pimentel2005}. For example, aphids
are well-known pests in numerous agricultural
contexts~\cite{Blackman2000, VanEmden2007, Dedryver2010}. However, in
cacao farms, the aphid \mbox{\textit{Toxoptera aurantii}} is not
perceived as a pest due to effective biological control by insect
predators~\cite{Blackman2000}. Aphids, specifically
\mbox{\textit{Toxoptera aurantii}}, are small sap-sucking insects that
can reproduce rapidly, potentially leading to large
infestations~\cite{Dixon1977}. These infestations can reduce crop
yields and quality, posing significant
farmer challenges~\cite{Oerke2006}.

Aphids not only feed on plant sap, weakening the plant, but they also
\mbox{excrete} honeydew, which can lead to the growth of sooty mold
and attract other pests~\cite{VanEmden2007}. In many environments, the
rapid reproduction and feeding behavior of aphids result in severe
agricultural losses~\cite{VanEmden1972}. Their ability to transmit
plant viruses further exacerbates their impact, making them one of the
most formidable pests in agriculture~\cite{Sylvester1989}. However,
intriguingly, in the cacao farms of \mbox{Ilheus}, Bahia, Brazil,
aphids do not always manifest as severe
pests~\cite{FirempongKumar1975, SilvaPerfecto2013}. This observation
suggests that there might be underlying ecological dynamics that
prevent aphids from reaching pest status in these
systems~\cite{Way2000}. One possible explanation lies in the
interactions \mbox{between} aphids and their natural predators, such
as \mbox{syrphid} \mbox{larvae} (\mbox{\textit{Syrphidae}}
family). \mbox{Syrphid} \mbox{larvae}, commonly known as
\mbox{hoverfly} \mbox{larvae}, are voracious predators of
aphids~\cite{Kindlmann1999}. They play a crucial role in biological
control by preying on aphids and thus regulating their
populations~\cite{Hagen1971}. The effectiveness of \mbox{syrphid}
larvae as biocontrol agents makes them significant in integrated pest
management strategies~\cite{VandermeerPerfecto2006}. Without
predators, aphid population growth was typically
logistic~\cite{Snyder2001, Dixon2000, Dixon1985}. However, the
voracious and early action of predators is responsible for pest
population control before its typical exponential growth phase
\cite{Hagen1971}.

Through field experiments~\cite{SilvaPerfecto2013}, differential
equations (ODE)~\cite{Colato1} and \linebreak continuous-time
computational simulations in this article, we study the \linebreak
\mbox{effective} control of aphids by hoverflies in cacao. To unravel
this \mbox{phenomenon}, we developed a simple mean-field predator-prey
model with \mbox{aging} dynamics~\cite{Gurtin1979, Cushing1982}. This
model, proposed by our research team, \linebreak \mbox{simulates} the
intricate interplay between aphids and their natural predators in
\mbox{cacao} ecosystems. Unlike traditional models such as the
Lotka-Volterra framework, our study introduces another perspective by
considering two distinct classes of aphid populations: juveniles,
which are incapable of \mbox{reproduction}, and adult females, capable
of asexual reproduction~\cite{Abrams1996, Mukherjee2014}. It allows us
to capture more accurately the life cycle and reproductive strategies
of aphids, providing a more realistic depiction of the ecological
dynamics~\cite{Meuthen2019, Jusufovski2020}.

The main hypothesis of this study is that the presence of syrphid
larvae effectively controls the aphid population, preventing them from
becoming pests in cacao cultivation. Our research question is: How do
the interactions between aphids and syrphid larvae influence the
population dynamics of aphids in cacao farms, and why does this
interaction prevent aphids from reaching pest status? Our approach
involves the formulation of a system of differential equations with
three rate equations, assuming a constant predator population. By
analyzing steady-state solutions and the stability of fixed points, we
aim to understand the conditions under which aphid populations are
kept in check. Additionally, we perform simulations on both complete
and random networks to study the phase transition from active to
absorbing states and
observe population dynamics differences. By delving into the
intricacies of predator-prey interactions, this study aims to provide
insights into why aphids may not always pose a threat as pests in
cacao cultivation. Understanding these dynamics can inform more
effective pest management strategies, contributing to sustainable
agricultural practices.

\section{Model}
\label{model}

Since the pioneering studies of Volterra and Lotka \cite{Murray2007,
  Hirsch1974}, the use of interacting populations in the prey-predator
framework (also applicable to parasite-host interactions or
consumption-resources) has generated a considerable amount of research
aiming to understand the dynamics of this type of ecological
interaction in different systems \cite{GodfrayHassell1987,
  ArditiGinzburg1989}. Two main approaches have been applied in such
studies: differential equations \cite{AbramsGinzburg2000} and
computational simulations \cite{Mitchell1998}.

Previous studies have demonstrated the importance of predator-prey
\linebreak \mbox{dynamics} in agricultural ecosystems. Firempong (1975
and 1976) and \linebreak \mbox{Kumar}~(1975) provided foundational
insights into these interactions in \mbox{cacao} plantations
\cite{Firempong1976, FirempongKumar1975}.  The mathematical modeling
of ecological interactions has been extensively developed over the
years. Turchin (2003) and \linebreak \mbox{Hassell}~(2000) offered
significant contributions to the theoretical frameworks of
\mbox{population} dynamics \cite{Turchin2003, Hassell2000}. In
particular, Hassell's work on the dynamics of arthropod predator-prey
systems provides a comprehensive understanding of these interactions
\cite{Hassell1978}.

In this study, we developed a model for the dynamics of the
interaction between the aphid \textit{T. aurantii} and syrphid larvae,
which were treated as effective biological control agents of this
aphid.
We propose a detailed mean-field predator-prey model incorporating
aging dynamics to investigate the population dynamics between aphids
and their natural predators in cacao farms. This model, based on the
works of Nascimento Silva~\cite{SilvaPerfecto2013}, simulates the
interaction between aphids (\textit{Toxoptera aurantii}) and syrphid
larvae (\mbox{\textit{Syrphidae}}), providing insights into why aphids
do not reach pest status in these ecosystems.

We assume that the predator population, syrphid larvae, remains
\mbox{constant} during the observation period. This assumption
simplifies the model and \mbox{focuses} on the prey dynamics, which
are influenced by predation and \linebreak \mbox{reproduction}
rates. The mean-field approximation is employed to reduce the
complexity of the spatial interactions, assuming that each individual
\mbox{interacts} with all others.

The key parameters of the model are:
\begin{itemize}
\item[-] $\alpha$: the rate at which juveniles mature into adults, representing the maturation process of aphids;
\item[-] $\beta$: the predation rate on juveniles, indicating the effectiveness of predators on young aphids;
\item[-] $\gamma$: the reproduction rate of adult aphids, reflecting their asexual reproductive capabilities;
\item[-] $\epsilon$: the predation rate on adult aphids, showing the impact of predation on mature aphids;
\item[-] $k_x$: the number of neighboring sites, representing the local interaction range.
\end{itemize}

The model consists of a system of differential equations that describe
the rate of change for juveniles~$J_x(t)$, adults~$A_x(t)$, and
the available space~$E_x(t)$ for the aphid population:
\begin{displaymath}
  \left\{
  \renewcommand{\arraystretch}{2.}
  \begin{array}{rcl}
    \displaystyle\frac{d J_x(t)}{dt} & = & -\alpha J_x(t) - \beta J_x(t) + \frac{\gamma}{k_x} \sum_y P_t[E_x(t),A_y(t)] \, , \\
    \displaystyle\frac{d A_x(t)}{dt} & = & \alpha J_x(t) - \epsilon A_x(t) \, , \\
    \displaystyle\frac{d E_x(t)}{dt} & = & \beta J_x(t) - \frac{\gamma}{k_x} \sum_y P_t[E_x(t),A_y(t)] + \epsilon A_x(t) \, ,
  \end{array}
  \right.
\end{displaymath}

where:
\begin{itemize}
\item[-] $J_x(t)$ represents the probability of there being a juvenile aphid at the site~$x$ at time instant~$t$;
\item[-] $A_x(t)$ represents the probability of there being an adult aphid at the site~$x$ at time instant~$t$;
\item[-] $E_x(t)$ represents the probability that the site~$x$ is empty and available for an aphid at time instant~$t$.
\item[-] $P_t[E_x(t),A_y(t)]$ is the joint probability that, at time
         instant~$t$, the site~$x$ is empty and available for an aphid
         and there is an adult aphid at the site~$y$.
\end{itemize}

We used numerical methods to solve the differential equations in the
\mbox{simple} mean-field case, employing algorithms such as the
fourth-order \linebreak Runge–Kutta method implemented in the C
programming language. The model was validated by comparing its
predictions with empirical data from cacao farms and with results from
other established models. Furthermore, we performed simulations on
complete graphs and random networks, which are expected to yield
results similar to those from \mbox{simple} mean-field calculations,
thus testing the self-consistency of the model. Additionally,
simulations on random networks explore the limitations of
\mbox{simple} mean-field predictions by considering the inherent
finite-size effects of the aphid population, providing a more
realistic understanding of the model's behavior.

\section{Mean-field results}
\label{mf}

In this section, we employ a single-site mean-field approximation to
\linebreak \mbox{simplify} the complex interactions of our
predator-prey model. Our goal is to analyze the population dynamics
and system stability under various \mbox{parameter} settings. By
focusing on average effects rather than detailed \mbox{spatial}
dynamics, we aim to derive insights into the overall behavior of the
model and identify conditions that prevent aphids from becoming
pests. The mean-field approximation assumes that each individual
\mbox{interacts} with the \mbox{entire} population, thus reducing the
system's complexity. This approach \mbox{allows} us to derive
tractable analytical expressions and gain insights into the overall
system behavior, which would be challenging to achieve with detailed
spatial models.

We begin by setting \mbox{$P_t[E_x(t),A_y(t)] \approx E_x(t)A_y(t)$}
and assuming homogeneous populations of juveniles, adults, and empty
spaces. Thus, we
use \mbox{$J_x(t)=J(t)$}, \mbox{$A_x(t)=A(t)$}, \mbox{$E_x(t)=E(t)$},
and \mbox{$A_y(t)=A(t)$} for all~$x$ and~$y$. With these assumptions,
we approximate the interactions as follows:
\begin{eqnarray*}
  P_t[E(t),A(t)] & \approx & E(t) A(t) \, , \\
  \sum_y P_t[E_x(t),A_y(t)] & \approx & k_x E(t) A(t) \, .
\end{eqnarray*}
This simplification leads to the following system of differential
equations describing the rates of change for juveniles, adults, and
empty spaces:
\begin{displaymath}
  \left\{
  \renewcommand{\arraystretch}{2.}
  \begin{array}{rcl}
    \displaystyle\frac{d J(t)}{dt} & = & -\alpha J(t) - \beta J(t) + \gamma E(t) A(t) \, , \\
    \displaystyle\frac{d A(t)}{dt} & = & \alpha J(t) - \epsilon A(t) \, , \\
    \displaystyle\frac{d E(t)}{dt} & = & \beta J(t) - \gamma E(t) A(t) + \epsilon A(t) \, .
  \end{array}
  \right.
\end{displaymath}


\subsection{Fixed points}
\label{fp}


Since solving this system of coupled differential equations
analytically is complex, we instead find the fixed points by setting
the derivatives to zero:
\begin{displaymath}
  \left\{
  \renewcommand{\arraystretch}{1.5}
  \begin{array}{rcl}
    0 & = & (-\alpha - \beta ) J^* + \gamma [1 - (J^* + A^*)] A^* \, , \\
    0 & = & \alpha J^* - \epsilon A^* \Rightarrow A^* = \alpha J^*/\epsilon .
  \end{array}
  \right.
\end{displaymath}
These equations yield two potential fixed points:
\begin{displaymath}
    \begin{array}{rcl}
    \left( J^*=0 \right. & \textrm{and} & \left. A^*=0 \right) \\
    & \textrm{or} & \\  
    \left\{
    J^* = \displaystyle\frac{[\gamma\alpha - \epsilon  (\alpha+\beta )]\epsilon }{\gamma\alpha(\epsilon  + \alpha)}
    \right.
    & \textrm{and} &  
    \left. A^*=\displaystyle\frac{\gamma\alpha - \epsilon  (\alpha+\beta )}{\gamma(\epsilon  + \alpha)} \right\}. 
    \end{array}    
\end{displaymath}

\subsection{Stability analysis of fixed points}
\label{saofp}


To determine the stability of these fixed points, we construct the
Jacobian matrix~$M$:
\begin{displaymath}
  M = \left[
    \renewcommand{\arraystretch}{1.5}
    \begin{array}{cc}
      -\alpha - \beta  - \gamma A^* & \gamma[1 - (J^* + 2A^*)] \\
      \alpha & -\epsilon 
    \end{array}
    \right] \, .
\end{displaymath}
By calculating the eigenvalues of this matrix, we can assess the
stability of each fixed point. The trivial fixed point $(J^*=0,
A^*=0)$ represents the absorbing state where the aphid population is
eradicated. This point loses stability when the components of the
non-trivial fixed point become positive, indicating a transcritical
bifurcation.

\subsection{Bifurcation and phase diagrams}%
\label{bapd}%

We observe that increasing the value of $\gamma$, the reproduction
rate of adult aphids, leads to a transcritical bifurcation at a
critical value $\gamma_{\textrm{c}}$, which \mbox{depends} on the
other parameters $\alpha$, $\beta$, and $\epsilon$. The absorbing
state (zero aphid population) remains a solution, but it loses
stability as the system transitions to a state with a finite
stationary population of aphids.%

\begin{figure}[!bht]
  \includegraphics[width=\columnwidth,angle=0]{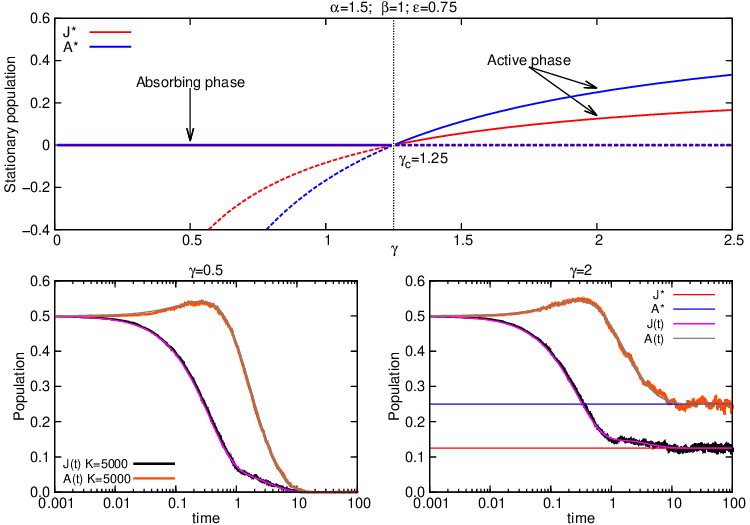}  
  \caption{Bifurcation diagram showing the transition from the
  absorbing state (no aphids) to a non-zero stationary state
  (persistent aphid population). The parameters are $\alpha=1.5$,
  $\beta=1$, and $\epsilon=0.75$. Below are two examples of the
  evolution of the normalized aphid population for each of the phases:
  on the left, for the absorbing phase; on the right, for the active
  phase. The insets zoom into the transition regions, highlighting the
  critical value $\gamma_{\textrm{c}}$ where the shift occurs. These
  detailed views are essential to understand the precise behavior of
  the system at critical points.}
\label{birfucation}
\end{figure}

\begin{figure}[!htb]
  \includegraphics[width=\columnwidth,angle=0]{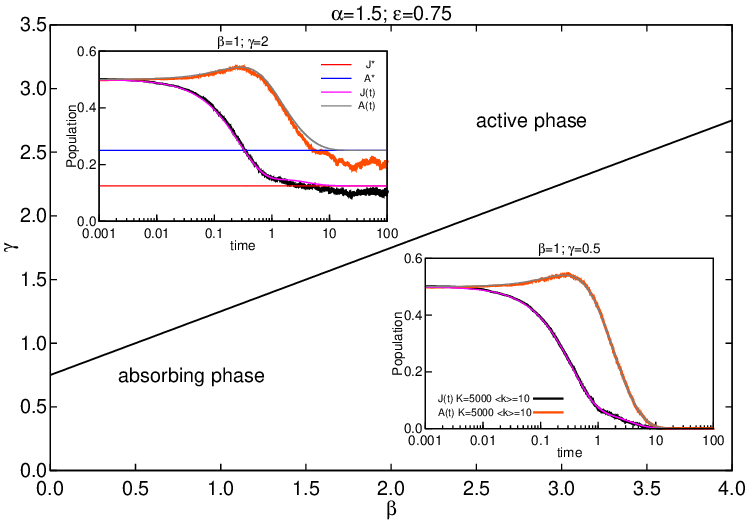}
  \caption{Phase diagram in the $\beta \times \gamma$ plane for fixed
  $\alpha=1.5$ and $\epsilon=0.75$. The diagram shows the regions
  where the system is in an active (non-zero aphid population) or
  absorbing (zero aphid population) state. In each inset, there is an
  example of the temporal evolution of the normalized aphid population
  corresponding to the phase where one is located in the diagram: on
  the top left, for the active phase; in the lower right, for the
  absorbing phase. The insets illustrate the different dynamic
  behaviors over time, clarifying the system's response to changes in
  $\beta$ and $\gamma$.}
  \label{phasediagrambetagamma}
\end{figure}

\begin{figure}[!htb]
        \centering
        \includegraphics[width=.8\linewidth]{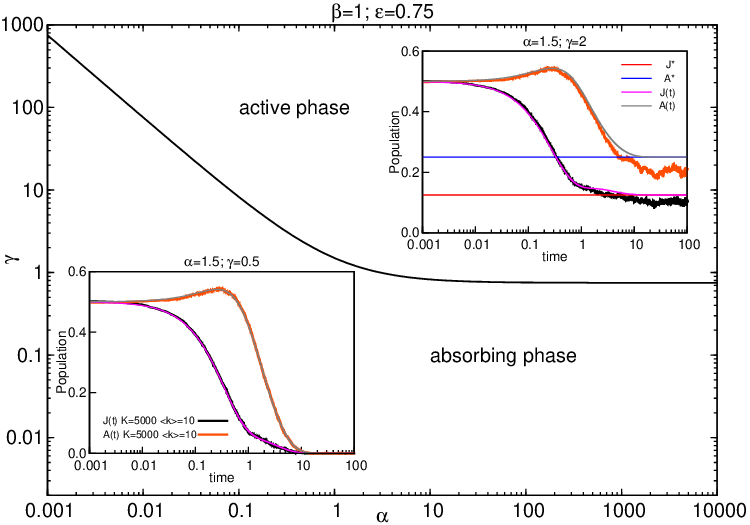}
        \caption{Phase diagram in the $\alpha \times \gamma$ plane for
        fixed $\beta=1$ and $\epsilon=0.75$. The diagram illustrates
        how variations in the rate at which juveniles become adults
        ($\alpha$) and the reproduction rate of adults ($\gamma$)
        affect the system's state. In each inset, there is an example
        of the temporal evolution of the normalized aphid population
        corresponding to the phase where one is located in the
        diagram: on the top right, for the active phase; in the lower
        left, for the absorbing phase. The insets provide detailed
        temporal evolution, highlighting the critical thresholds that
        influence the system's dynamics.}
        \label{phasediagramalphagamma}
\end{figure}

To visualize these transitions, we construct bifurcation and phase
diagrams. The bifurcation diagram (Figure \ref{birfucation}) shows the
transition from the absorbing to a non-zero stationary state as
$\gamma$ increases. The phase diagrams (Figures
\ref{phasediagrambetagamma} and \ref{phasediagramalphagamma}) in the
$\beta \times \gamma$ and $\alpha \times \gamma$ planes illustrate the
regions of parameter space where the system exhibits absorbing or
active phases.%

In the bifurcation diagram, the x-axis represents the reproduction
rate $\gamma$, and the y-axis shows the normalized aphid
population. As $\gamma$ crosses the critical value
$\gamma_{\textrm{c}}$, the system shifts from an absorbing phase (no
aphids) to an active phase (persistent aphid population). This diagram
provides a clear visualization of how the reproduction rate of adult
aphids influences the overall population dynamics, indicating a
critical threshold beyond which the aphid population becomes
sustainable.%

The phase diagrams further elucidate how changes in $\beta$ (predation
rate on juveniles) and $\alpha$ (the rate at which juveniles mature)
influence the system's behavior. The results are presented in
graphical form to illustrate the key findings of our analysis. Each
graph includes insets to highlight specific aspects of the data that
are crucial for understanding the overall results.%

\textbf{Figure \ref{birfucation}} shows the bifurcation diagram with the
x-axis representing the reproduction rate $\gamma$, and the y-axis
showing the normalized aphid population. The insets in the graphs
provide a zoomed-in view of critical regions that exhibit interesting
or significant behavior, such as the transition from the absorbing
state to a non-zero stationary state. These insets are necessary to
understand the finer nuances of the model's behavior and provide
additional context to the main findings.

\textbf{Figure \ref{phasediagrambetagamma}} shows the phase diagram in the
$\beta \times \gamma$ plane for fixed $\alpha=1.5$ and
$\epsilon=0.75$. The diagram delineates the regions where the system
is in an active (non-zero aphid population) or absorbing (zero aphid
population) state. The insets illustrate the temporal evolution of the
normalized aphid population corresponding to each phase, providing
clear examples of how the populations behave over time in different
parameter regimes.

\textbf{Figure \ref{phasediagramalphagamma}} illustrates the phase
diagram in the $\alpha \times \gamma$ plane for fixed \mbox{$\beta=1$}
and $\epsilon=0.75$. It shows how variations in the rate at which
juveniles \mbox{become} adults ($\alpha$) and the reproduction rate of
adults ($\gamma$) affect the system's state. The insets provide
detailed views of the temporal evolution in different \mbox{regions}
of the phase diagram, helping to understand the dynamic transitions
between phases.

\section{Simulation Results}
\label{sr}


We conducted extensive simulations on complete graphs and random
networks to complement our mean-field analysis. These simulations
verify \mbox{theoretical} predictions and offer nuanced insights into
the system's dynamics.%

\subsection{Complete Graph Simulations}%
\label{cgs}%

In simulations on complete graphs, every node is connected to every
other node, which mirrors the mean-field assumption of uniform
interactions. The results align well with the mean-field predictions,
especially in the absorbing phase where the aphid population is
zero. In the active phase, the simulated populations of juveniles and
adults closely match the steady-state solutions predicted by the
mean-field model. This consistency underscores the robustness of the
mean-field approximation for large, well-mixed populations.%

\subsection{Random Network Simulations}%
\label{rns}%

Simulations on random networks, where connections between nodes are
assigned randomly, show behaviors that complement the mean-field
predictions. While the general trend of the phase transition from
absorbing (no aphids) to active (persistent aphid population) states
is consistent with the mean-field analysis, the exact parameter values
at which these transitions \mbox{occur} can vary due to
network-induced fluctuations and finite size effects. The transition
points observed in simulations on random networks broadly align with
those identified in the mean-field analysis. However, minor
differences can occur due to the inherent variability in network
structures.

Additionally, these simulations exhibit variability in exact
population \mbox{levels} at equilibrium relative to mean-field
predictions, highlighting the \linebreak \mbox{influence} of network
topology on the dynamics. Specifically, the population levels at
equilibrium in the simulations do not exactly match those predicted by
the mean-field model. There is noticeable variability in these
equilibrium population levels, meaning that the equilibrium population
can fluctuate or differ from the mean-field predictions due to factors
such as network topology, random interactions, or other stochastic
effects inherent in the simulations.

By systematically varying parameters $\alpha$, $\beta$, $\gamma$, and
$\epsilon$ in the simulations, we mapped out the phase transitions in
the system. The critical values at which the system transitions from
the absorbing to the active phase are consistent with the bifurcation
points identified in the mean-field analysis. However, the simulations
demonstrate that random networks can sustain larger fluctuations.

\subsection{Discussion}%
\label{discussion}%

The simulation results provide valuable insights into the practical
implications of predator-prey dynamics in agricultural settings. The
ability of \mbox{syrphid} larvae to control aphid populations is
confirmed, but the effectiveness of this biological control strategy
can vary based on the spatial distribution and connectivity of the
habitat. These findings suggest that enhancing habitat connectivity
and promoting a well-mixed predator-prey environment could improve the
stability and effectiveness of natural pest control mechanisms.

Overall, the simulation results reinforce the theoretical findings of
the mean-field analysis while also uncovering additional layers of
complexity \mbox{introduced} by network structure and finite-size
effects. These insights are crucial for developing more nuanced and
effective strategies for ecological management and pest control in
agricultural ecosystems.

\section{Conclusions}
\label{conclusions}

In this study, we investigated the predator-prey dynamics between
aphids (\textit{Toxoptera aurantii}) and syrphid larvae
(\textit{Syrphidae}) in cacao farms using a mean-field approximation
and extensive simulations. Our goal was to understand why aphids do
not reach pest status in these ecosystems and how natural predation
controls their population.

Our model extends classical predator-prey frameworks by incorporating
key biological factors such as the maturation process of aphids and
their asexual reproduction, offering a stage-structured
perspective. By distinguishing between juvenile and adult aphids, we
capture more nuanced population dynamics, addressing gaps in
traditional models that lack life-stage
differentiation~\cite{DostalkovaKindlmannDixon2002, Gurtin1979,
  Cushing1982, Abrams1996, Murray2007}.

We derived steady-state solutions and analyzed the stability of the
system's fixed points. The results showed that while the absorbing
state (zero aphid population) is possible, it is not always stable. A
nonzero stationary solution emerges with an appropriate set of
parameters, confirmed through phase diagrams.

Simulations on complete graphs validated our mean-field predictions,
demonstrating that the model accurately represents the system's
dynamics under well-mixed conditions.  However, simulations on random
networks revealed additional complexities, such as larger fluctuations
in population levels at equilibrium, which were not fully predicted by
the mean-field approach. These results highlight the significant role
that network topology and stochastic effects play in population
stability, particularly in pest management contexts~\cite{Turchin2003,
  Hanski1999}.

Our findings contribute to theoretical biology by showing that the
interplay between life-stage structure, predation, and spatial
heterogeneity is crucial for understanding how predator-prey systems
remain stable. These results underscore the importance of network
effects in population dynamics and provide a more refined theoretical
framework for biological pest control in agricultural ecosystems.

From a practical perspective, the study reinforces the effectiveness
of natural biological control mechanisms. The presence of syrphid
larvae controls aphid populations, preventing them from reaching pest
status. Enhancing habitat connectivity and promoting a well-mixed
predator-prey environment could further improve the stability and
effectiveness of pest control strategies \cite{Altieri1995,
  Tscharntke2007}.

Future research should explore the influence of different network
topologies on predator-prey dynamics and how spatial structure impacts
population stability.  Empirical studies are needed to validate these
theoretical and simulation results in real-world agricultural
settings. Additionally, further work should investigate how
interactions between multiple predator and prey species, as well as
environmental factors such as climate change, affect these dynamics
\cite{Lipsitch95}. In conclusion, this study integrates concepts from
predator-prey dynamics, network theory, and biological control,
contributing to a deeper understanding of ecological interactions in
agroecosystems and offering insights that can guide sustainable pest
management practices \cite{VandermeerPerfecto2006}.

\section*{Acknowledgements}
V.R.V.A., N.G.F.M., E.N.S. and A.T.C.S. thank the Department of
\linebreak \mbox{Education} of the government of the State of Bahia
for the incentive bonus for scientific production. E.N.S. thanks CAPES
Foundation-Brazil for a scholarship and grants that allowed data
collection in the field (BEX 1609/00-9). V.R.V.A. thanks Professor
Dagoberto da Silva Freitas for encouraging him to return to scientific
research. A.C. thanks the Centro de Ciências Agrárias of UFSCar for its
support of scientific research.


\bibliographystyle{unsrt}
\bibliography{vladimir}

\end{document}